\documentclass[runningheads]{llncs}
\usepackage[T1]{fontenc} 
\usepackage[utf8]{inputenc} 
\usepackage{amsmath}
\usepackage{url}
\usepackage{graphicx}
\usepackage[caption=false]{subfig}
\usepackage{color}
\usepackage{xcolor,colortbl}

\usepackage{svg}
\usepackage{float}

\usepackage{adjustbox}
\usepackage{makecell}
\usepackage{booktabs}
\usepackage{natbib}
\usepackage[inline]{enumitem}
\usepackage{nicefrac}
\usepackage{todonotes}

\graphicspath{{img/}}

\providecommand*{\quarternote}{%
  \begingroup
    \fontencoding{U}%
    \fontfamily{wasy}%
    \selectfont
    \symbol{12}%
  \endgroup
}

\begin{document}

\makeatletter 
\let\c@table\c@figure
\let\c@lstlisting\c@figure
\let\ftype@table\ftype@figure
\let\ftype@listings\ftype@figure
\makeatother

\title{Medley2K: A Dataset of Medley Transitions}

\author{Lukas Faber$^\star$ \and Sandro Luck$^\star$ \and Damian Pascual$^\star$ \and Andreas Roth\thanks{Authors in alphabetical order.} \and Gino Brunner \and Roger Wattenhofer}
\authorrunning{L. Faber et al.}
%
\institute{ETH Zurich, Switzerland\\
\email{\{lfaber,dpascual,brunnegi,wattenhofer\}@ethz.ch} \\ \email{\{sluck,rothand\}@student.ethz.ch}}

\maketitle

\begin{abstract}
The automatic generation of medleys, i.e., musical pieces formed by different songs concatenated via smooth transitions, is not well studied in the current literature. To facilitate research on this topic, we make available a dataset called Medley2K that consists of $2,000$ medleys and $7,712$ labeled transitions. Our dataset features a rich variety of song transitions across different music genres. We provide a detailed description of this dataset and validate it by training a state-of-the-art generative model in the task of generating transitions between songs.

\end{abstract}

\section{Introduction}

Automatic music generation has undergone major development in the last few years, thanks to the progress of deep learning. Indeed, previous studies have demonstrated the ability of deep learning models to generate pleasant music in many different applications~\cite{yang2017midinet,  dong2018convolutional, oord2016wavenet, waite2016magenta, 2019InpaintNet, lewandowski2012modeling, briot2019musicbook,  brunner2018midivae, huang2019music, mogren2016crnngan}. 
To perform well, these models need large amounts of training data and, depending on the application, collecting such data may not be a trivial task. In this work, we contribute to the growing field of automatic music generation by presenting Medley2K, a new dataset for MIDI medley composition.


A medley is a special type of music piece that is formed by connecting different songs through specifically crafted musical transitions. Despite the popularity of medleys, existing literature has not addressed transition generation, yet. One reason for this is the lack of medley-specific datasets, which hampers progress in this field. Collecting such a dataset is challenging since it requires precise annotations of the transitions between individual songs. 
Our dataset contains machine-readable labeled transitions extracted from $2000$ human-curated medleys. In this work, we give a complete description of Medley2K, and we empirically show that it can be used to train deep generative models for medley transition generation.




\section{Related Work}
Although a considerably large number of datasets for music modeling are publicly available (a sample collection can be found here\footnote{\url{https://ismir.net/resources/datasets}}), none of those datasets is tailored to medley composition. In particular, the name-related MedleyDB dataset \citep{bittner2014medleydb, bittner2016medleydb} contains polytrack music of single songs rather than medleys. Conversely, our dataset consists of medley pieces with detailed labels on the transition points in order to foster further work on automatic medley composition.



\section{Dataset: Medley2K}

\begin{figure}[htp]
\centering
\subfloat[Instrumentation distribution. Every bar corresponds to one instrument; the first bar is ``Acoustic Grand Piano''.\label{fig:dataset_instrumentation}]{%
  \includegraphics[width=0.45\textwidth]{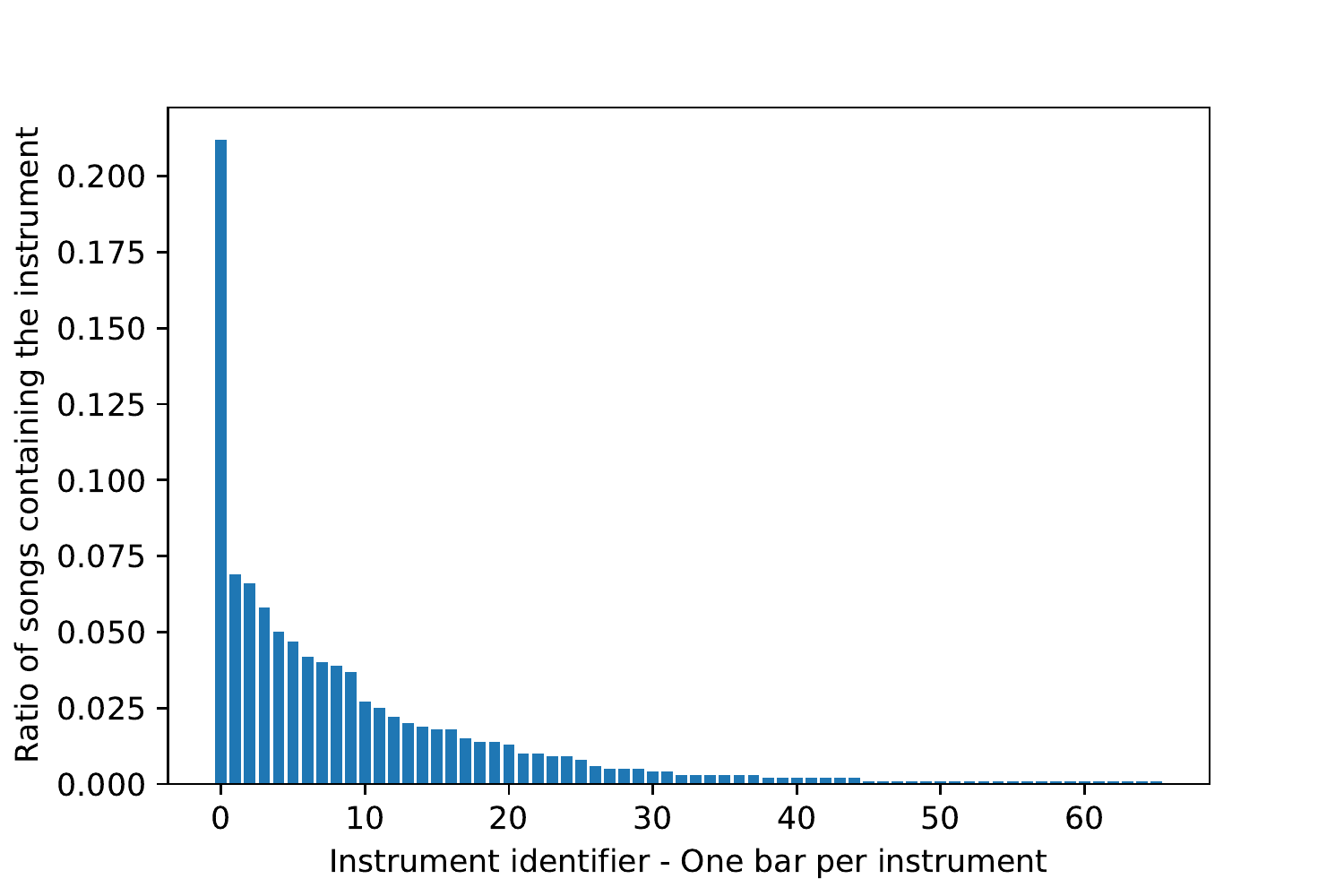}
}\hfil
\subfloat[Frequencies of notes played after postprocessing. Every bar corresponds to one number.\label{fig:transition_notes}]{%
  \includegraphics[width=0.45\textwidth]{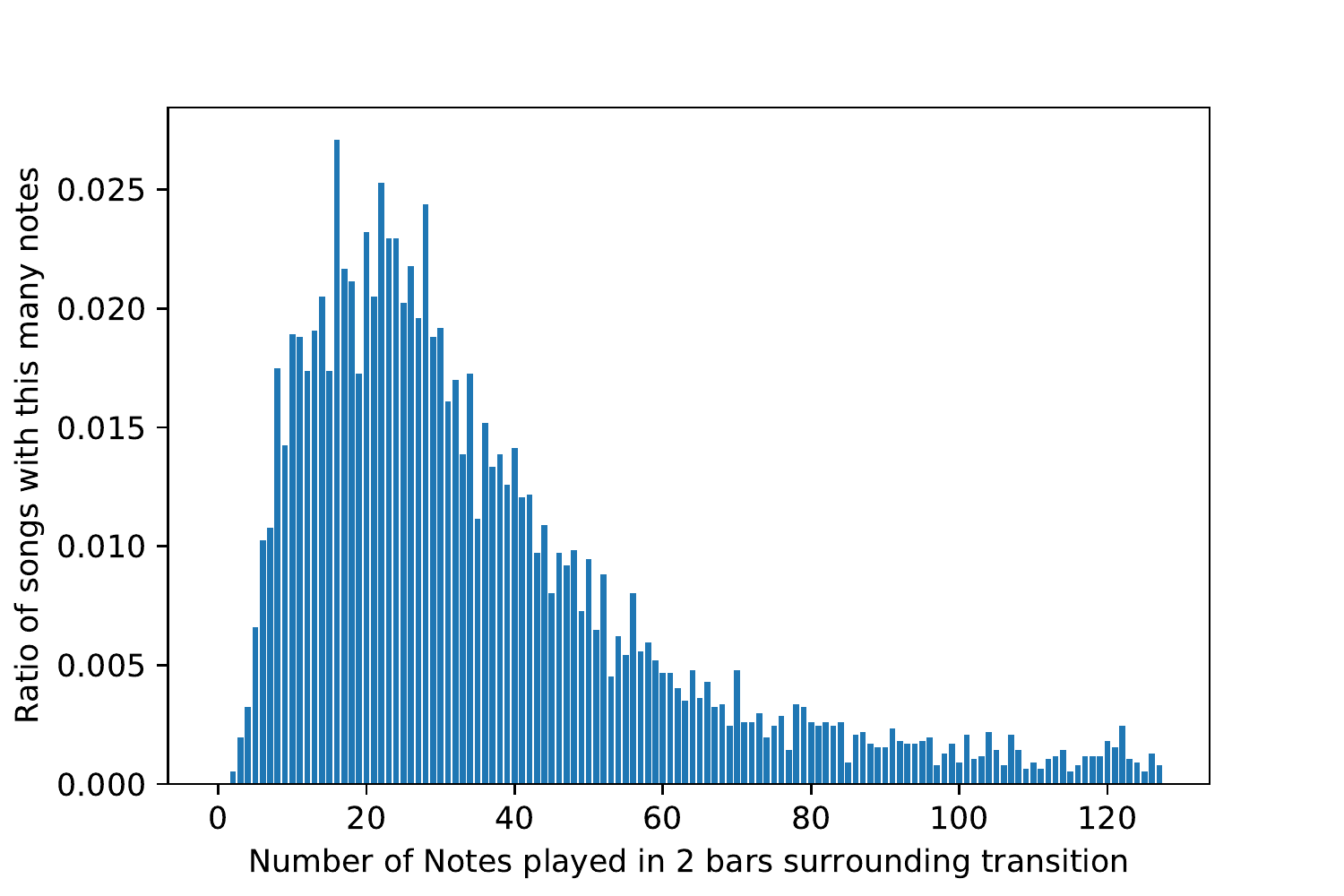}%
}

\label{fig:image2}

\end{figure}

Medley2K is a new dataset that consists of $2000$ human-created medleys crawled from the website \url{musescore.com} with a total of $10,269$ transitions. All medleys are licensed as shareable, while only a subset is available for commercial use. The dataset contains a rich variety of medleys spanning across several paces, musical scales, and genres. The medleys are on average $6$ minutes long with $17.47$ key changes and $9.99$ tempo changes. Furthermore, the medleys in the dataset have rich instrumentation, featuring an average of $7.65$ different instruments. Additionally, as seen in Figure~\ref{fig:dataset_instrumentation}, all instruments except for the ``Acoustic Grand Piano'' occur in less than $10\%$ of medleys, which means that the instrumentation varies largely across samples.
The medleys from musecore come as MIDI files, together with PDF scoresheets and a machine-readable MXL file.
Typically, composers annotate the point in time where one song in the medley transitions into the next in the scoresheet. These transitions usually start at the beginning of a new bar. We parse the MXL file for annotation indicating such transitions points. To ensure data quality, we filter annotations that do not indicate a transition; in particular, we ignore annotations of numbers, musical symbols (such as \quarternote), or a manually defined blacklist of musical expressions (such as ``vivante'').

We evaluate the quality of this extraction method on $30$ medleys manually labeled, with a total of $205$ actual transitions. Table~\ref{tab:tp_extraction} shows the confusion matrix between the actual transition points and the labels given by our extraction method. Overall, the automated extraction achieves a precision of $90.70\%$ and a recall of $57.07\%$. Note that the high precision value indicates that what we identify as a transition (and will potentially feed into a machine learning model) is very likely a genuine transition. The recall means that we can still extract more transition points, i.e., assuming the method has a similar recall over the whole dataset we could find around $18,000$ transitions. Thus, the labeling --- while not complete --- is of high quality.

\begin{table}[H]
\centering
\begin{tabular}{lcc}
\toprule
& True Positive &  True Negative\\\midrule
Predicted Positive & 117 & 12 \\\midrule
Predicted Negative & 88 & 4370 \\
\bottomrule
\end{tabular}
\caption{Validation of Labeling Process}
\label{tab:tp_extraction}
\end{table}

\vspace{-0.3cm}

Next, we examine the notes around the transition point. We observe that for a large fraction of transitions (around $30\%$) the music around the transition point contains only silence or a single long-held note. These samples cannot be used to learn transitions that consist of more than one note. Since we want to focus on musically pleasant transitions with different notes, we filter the data by looking at the two half bars preceding and the two half bars following the transition point. If a new note starts in either of those four half bars, we keep the transition, otherwise we discard it. This way, each transition consists of at least four played notes. 
In Figure~\ref{fig:transition_notes} we show in more detail the number of notes played. The filtered transitions have a high variety of notes ranging from four to more than $60$ notes. 
After this postprocessing, we compose the final dataset with a total of $7,712$ labeled transitions.

\vspace{-0.3cm}


\vspace{-0.5cm}

\section{Experimental Evaluation}
In this section, we conduct an experimental study on the validity of the Medley2K dataset for automatic medley composition. 
To this end, we use a deep neural model that learns to generate Medley transitions as a specialization of the task of filling gaps in music --- also called music inpainting. We build on the InpaintNet architecture by \citet{2019InpaintNet} and extend it to support polyphonic music while keeping the same hyperparameters. Given that some internal components of the InpaintNet architecture are tailored to 4/4 beats, we omit in this experiment all transitions with a different beat, resulting in $4,662$ transition points. For each transition, we generate a sample by taking the four bars around the label plus the four bars preceding (past context) and following (future context) the transition point, i.e., 12 bars in total. We encode the data from our Medley2K dataset with a scheme similar to \citet{hadjeres2017deepbach}, except that instead of using one symbol for holding the previous note, we use one extra symbol \emph{per note} to denote ``Hold''. Although it doubles the number of classes, we found that this encoding reduces class imbalance and improves model performance.  

\begin{figure}
    \centering
    \includegraphics[width=0.5\textwidth]{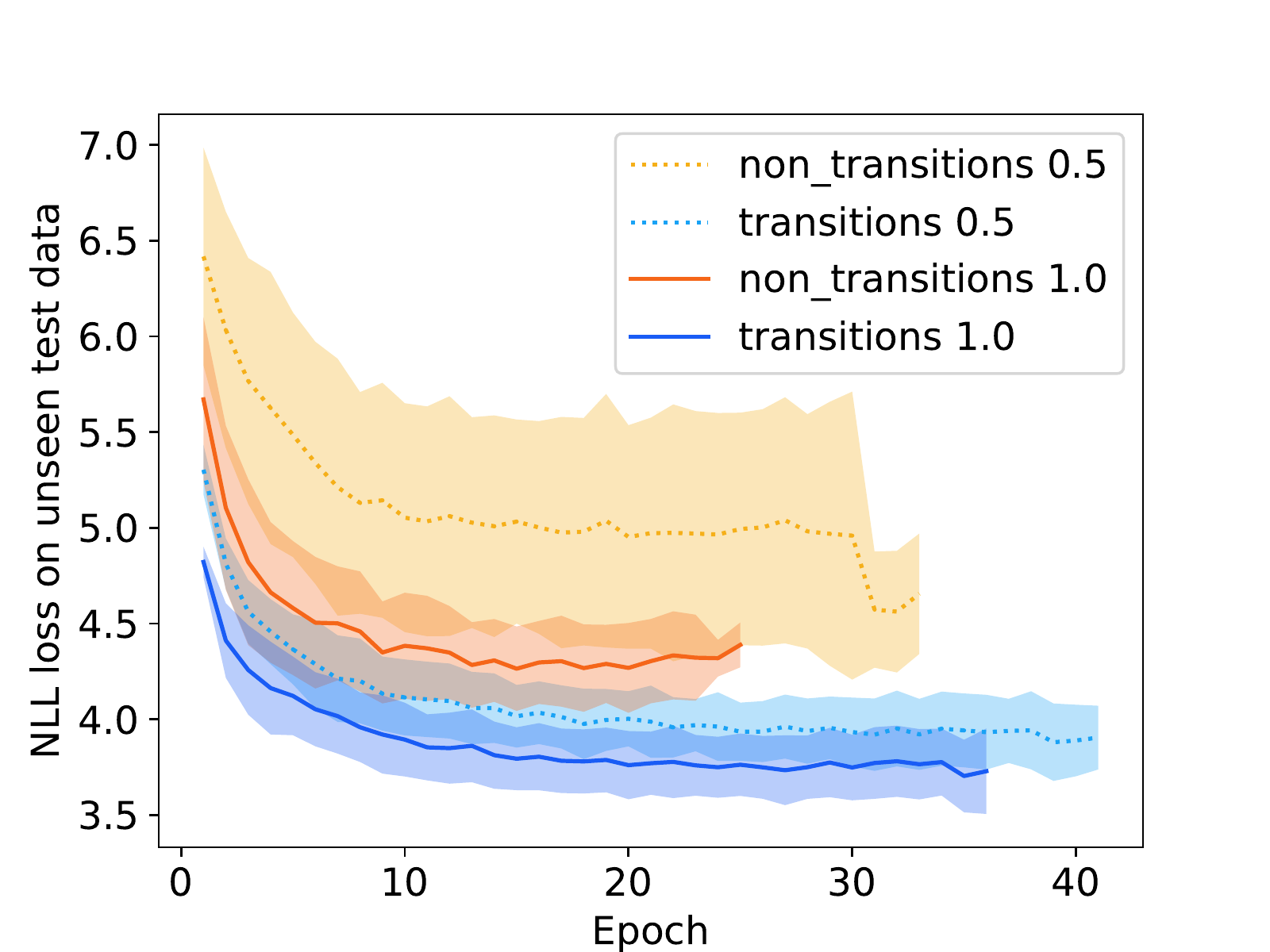}
    \caption{Performance (NLL loss) of the generative model. Solid lines denote full training sets, dotted lines half training sets. Training on transitions only (bottom lines) achieves better results than training on general music (top lines).}
    \label{fig_dataset_eval}
\end{figure}

To validate our dataset, we compare two models, one trained with transition data and one trained with arbitrary portions of music from the dataset. We split the transition data into $80/10/10$ for training, validation, and test, where the test data is used to evaluate both models. Furthermore, for each model, we consider two training sets, one consisting of 100\% of the training transitions (or the equivalent number of samples of arbitrary music), and one with 50\% of the samples.
The results of these experiments are shown in Figure~\ref{fig_dataset_eval}, which shows that given the same amount of data, training on transition data yields better performance on the test set than arbitrary music.
In fact, even using only 50\% of the transition training is better than using twice as much data of arbitrary music, which demonstrates that training on transitions largely benefits the automatic composition of medleys. This validates our Medley2K dataset as a valuable tool for further work in automatic medley generation.

\vspace{-0.3cm}

\vspace{-0.5cm}

\section{Conclusion}
We make available\footnote{\url{https://polybox.ethz.ch/index.php/s/STSczoZ2e0IcoVf}} the first dataset for medley composition of MIDI music. The dataset has a rich variety of music pieces, instrumentation, key changes, and tempos. We provide machine-readable labels for $7,712$ transition points and validate the dataset by demonstrating its ability to train a state-of-the-art model for music generation. We expect that this dataset will encourage further research in the field of medley generation and automatic medley detection.

\bibliographystyle{splncsnat}
\bibliography{bibfile}

\end{document}